\documentclass{IEEEtran}
\usepackage[hmargin=1in,vmargin=1in]{geometry}   
\usepackage{cite}   
\usepackage{pgf}
\usepackage{tikz}
\usepackage[cmex10]{amsmath}
\usepackage{amssymb}
\usepackage{url}
\usepackage{pbox}
\usepackage{setspace}
\usepackage{wrapfig}
\usepackage{multirow}
\usepackage{xspace}
\usepackage{array}
\usepackage{hyperref}
\usepackage{booktabs}

\usetikzlibrary{arrows,shapes,positioning,calc,fit,backgrounds,decorations.markings,trees}
\pgfdeclarelayer{background}
\pgfdeclarelayer{foreground}
\pgfsetlayers{background,main,foreground}

\tikzset{>=latex}
\tikzset{
  smallarrow/.style={
    decoration={markings,mark=at position 1 with {\arrow[scale=0.8]{>}}},
    postaction={decorate},
    shorten >=1.5pt
}}
\tikzset{
  smallarrowdub/.style={
    decoration=
    {
      markings,
      mark=at position 0.023 with {\arrow[scale=0.8]{<}},
      mark=at position 1 with {\arrow[scale=0.8]{>}}
    },
    postaction={decorate},
    shorten >=1.5pt,
    shorten <=1.5pt
}}

\def\authnotes{0}

\newcounter{notectr}[section]
\newcommand{\thenote}{\thesubsection.\arabic{notectr}\refstepcounter{notectr}}

\newcommand{\note}[2]{$\ll$#1~\thenote: #2$\gg$}
\newcommand{\tnote}[1]{\ifnum\authnotes=1 \textcolor{blue}{\note{TomR}{#1}}\fi}

\newcommand{\fixme}[1]{\ifnum\authnotes=1{\textcolor{red}{[FIXME: #1]}}\fi}
\newcommand{\tsnote}[1]{\ifnum\authnotes=1 \textcolor{magenta}{\note{TomS}{#1}}\fi}

\newcommand{\figref}[1]{Fig.~\ref{#1}}

\newcommand{\subparagraph}[1]{\paragraph{#1}}
\renewcommand{\paragraph}[1]{\vspace*{3pt}\noindent\textbf{#1}}

\begin{document}

\title{Network Traffic Obfuscation and\\ Automated Internet Censorship}

\author{Lucas Dixon \hspace*{3em} Thomas Ristenpart \hspace*{3em}
Thomas Shrimpton \\ Jigsaw, Cornell Tech, University of Florida}

\maketitle

\thispagestyle{plain}
\pagestyle{plain}

%

\begin{abstract}
Internet censors seek ways to identify and block internet access to information
they deem objectionable.  Increasingly, censors deploy advanced networking tools
such as deep-packet inspection (DPI) to identify such connections. In response,
activists and academic researchers have developed and deployed network traffic
obfuscation mechanisms. These apply specialized cryptographic tools
to attempt to hide from DPI the true nature and content of connections.

In this survey, we give an overview of network traffic obfuscation and its role
in circumventing Internet censorship. We provide historical and technical
background that motivates the need for obfuscation tools, and give an overview
of approaches to obfuscation used by state of the art tools.  We discuss the
latest research on how censors might detect these efforts. We also describe the
current challenges to censorship circumvention research and identify concrete
ways for the community to address these challenges.
\end{abstract}

%

\section{Introduction}
\label{sec:intro}
Around the world, well-resourced governments seek to censor the Internet.
Censorship is driven by the desires of governments to control access to information deemed
politically or socially sensitive, or to protect national economic
interests. For example, in China
performing Internet searches for information on Tiananmen square reveals no
information about the events in 1989, and communication platforms run by
companies outside China, such as GMail and Twitter, are among those that
are routinely blocked. Although China's ``Great Firewall'' is perhaps the best known example of Internet censorship by a nation-state, similar controls are enacted in Turkey, Iran, Malaysia, Syria, Belarus
and many other countries.\footnote{An excellent survey of the censorship policies by various governments around the world
is provided by the OpenNet Initiative \url{https://opennet.net/}.}

In this article we will focus on the technical underpinnings of \emph{automated
censorship} and of the \emph{network traffic obfuscation tools} aimed at
circumventing it. By automated censorship, we mean government
deployed or mandated network monitoring equipment that is typically installed  at Internet service providers (ISPs).  These systems detect and disrupt the flow of censored information
without any direct human involvement and are applied broadly to enforce
government policies over all citizens.  This
differs from targeted approaches to taking down Internet content, such as
the United
States government's take-down of the Silk Road, which was an Internet
marketplace for illicit goods such as drugs and weapons.

We will first review how modern deep-packet inspection (DPI) devices
enable censors to filter traffic based on the application-layer content they
contain, e.g., search terms, requested URLs, filetypes, and information that
identifies the application itself.
We will then explore the technical methods used to obfuscate application-layer
content, in order to minimize the usefulness of DPI.  These make intimate use of
cryptography, though in non-standard ways. We place the methods into one of
three categories: those that randomize every bit of packet payloads, those that
transform payloads so that they mimic typical ``benign'' traffic (e.g., HTTP
messages), and those that tunnel payloads through actual implementations of
network protocols (e.g., a VPN or HTTPS proxy).  Each category comes with its
own set of assumptions about the censor's approach, and targets a different
point in the security/performance design space.

We then discuss the challenges to progress faced by researchers in this
area.  Chief among these is the dearth of solid information about the technical
capabilities of real-world censors, and their sensitivities to filtering errors
(both missing targeted traffic and blocking non-targeted traffic).  Similarly,
we lack understanding about real-world users of censorship-circumvention tools.
In both cases, even gathering such information presents significant challenges,
as we will see.

The final contribution of this article is to suggest
concrete tasks and research efforts that would give considerable aid to the
design of future obfuscation tools.

%

\section{Background: From Censorship to\\ Protocol Obfuscation}
\label{sec:background}

In what follows, we will use terms like ``censor'', ``adversary'' and ``sensitive flow'' repeatedly.  To aid understanding,
we provide a summary of such key terms in \figref{fig:terms}.

The technical means of automated censorship assumes that the censor enjoys privileged access to network infrastructure.
Governments typically have this via direct or indirect control of Internet service providers (ISPs).
In the security community this is called an on-path adversary: it is on the path of network packets between the
user seeking to access sensitive content, and the provider of that content (e.g., a web server).

\begin{figure}[t]
\center{
\footnotesize
\renewcommand{\arraystretch}{1.5}
\begin{tabular}{>{\flushleft}m{0.6in}  >{\flushleft\arraybackslash}m{2.2in}}
\toprule
{\bf Sensitive flow} & Network communications targeted for being offensive, counter-regime, or using obfuscation tools\\
{\bf Censor, Adversary} & The party that controls the communication network and is attempting to identify and restrict sensitive flows of information \\
{\bf Obfuscation tool} & Software designed to prevent censors from identifying/blocking communications\\
{\bf Background traffic} & Non-sensitive flows coexisting (potentially) with sensitive flows on the network\\
{\bf Collateral damage} & Background traffic adversely affected by censor's efforts to control sensitive flows\\
{\bf Proxy} & Party who will forward traffic to broader Internet on behalf of censored client; external to censor-controlled network\\
\bottomrule
\end{tabular}
}
\caption{Summary of often-used technical terms.}
\label{fig:terms}
\end{figure}

\subsection{Traffic Identification Methods} Network censors face two main tasks:
to accurately identify network traffic that carries or attempts to access
sensitive content, and (potentially) to perform a censoring action on that
traffic.  Examples of the latter are dropping packets to degrade the user's
experience, forcing one or both of the endpoints to drop the connection (e.g. by
sending TCP reset messages), and redirecting web connections to a
censor-controlled web server.  But as censoring actions must be preceded by
identification of targeted traffic, we focus only on the censor's identification
task.

\paragraph{Address-based identification. } Originally, attempts
to access censored Internet content were identified by first associating
addressing information --- IP addresses and port numbers, domain names --- with
sensitive content, resulting in a blacklist of addresses.  Any network
connections to these addresses were deemed as attempts to access sensitive
content.  For blacklisted domain names,  the censor can
direct ISPs to run modified DNS software. Connections to these domains are then typically blocked or
misdirected.  For IP addresses, the censor can install hardware mechanisms at
ISPs that compare information in a packet's IP header against the list.  As the
IP headers appear early in the packet, one needs only to perform ``shallow''
packet inspection, disregarding information encapsulated deeper in the packet.
TCP or UDP port information is similarly available via shallow inspection.

\begin{figure}[t]
\center
\includegraphics[width=2.5in]{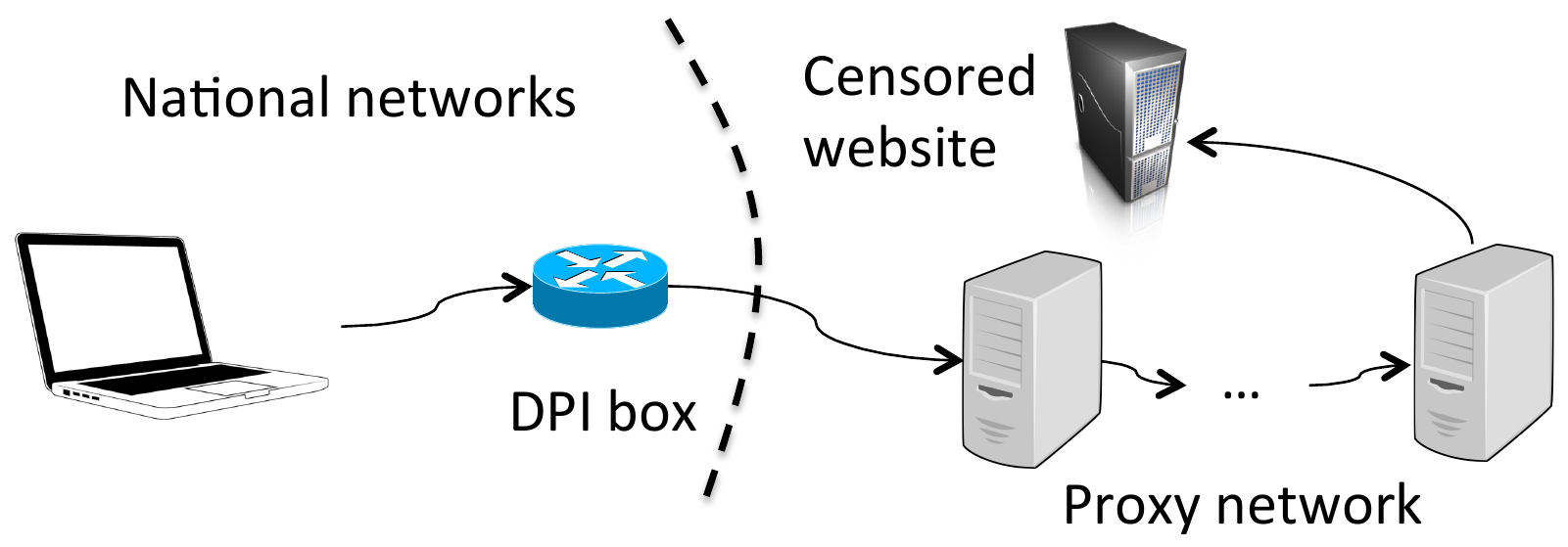}
\caption{Proxy-based circumvention of censorship. A client connects to a proxy
server that relays its traffic to the intended destination. It may send the
traffic through multiple proxies before the destination should anonymity be
desired.}
\label{fig:diagram-proxy}
\end{figure}

A user can avoid both domain-name- and IP/port-based identification by using a
\emph{proxy}, a cooperative machine whose address information is not
blacklisted. The operation of proxy-based circumvention is summarized by the diagram in~\figref{fig:diagram-proxy}. Many tools such as Tor,
uProxy, Anonymizer, and Psiphon rely on the use of proxies.\footnote{These tools are available online at: \url{https://torproject.org}, \url{https://www.uproxy.org}, \url{https://anonymizer.com}, \url{https://psiphon3.com}}
Recently, domain fronting~\cite{fifield2015blocking}, a form of decoy routing~\cite{decoyrouting,cirripede,telex} that
operates at the application layer, has also been successfully deployed and is discussed further in Sec.~\ref{sec:systems}.

\paragraph{Identification via deep-packet inspection.} The success of proxy
systems in practice has led censors to deploy new DPI mechanisms that
identify traffic based on information deeper inside the network packets.
For example, application-layer content carried in (unencrypted) packet payloads
can divulge user-generated data, such as keywords within search URLs. We already noted China's Great Firewall blocking traffic that contains blacklisted keywords.

In reaction to DPI, modern circumvention tools systems encrypt content to proxies, preventing this kind of keyword search. The result has been that sophisticated modern censors now use DPI to attempt to identify and block traffic created by circumvention tools. For example, China used DPI to detect Tor connections by looking for a specific
sequence of bytes in the first application-layer message from client to server.
This sequence corresponds to the TLS cipher suites that were selected by Tor
clients, which were unlike those selected by other TLS  implementations.
As another example,
Iran used DPI to determine the expiration date of TLS certificates.  At the
time, Tor servers used certificates with relatively near expiration dates, while
most commercial TLS servers chose expiration dates that are many years away.

These blocking policies used fairly sophisticated information about the circumvention tool
---the particular implementation of TLS--- being carried in packet
payloads.  But even a crude scan of application-layer content can discover that
standard protocols (e.g. SSH, TLS) are being transported.  This may be enough to
trigger insensitive blocking, or logging of source/destination IP addresses for
more detailed analysis or later action.

\subsection{Network Traffic Obfuscation}
The use of DPI to perform tool
identification motivated the circumvention-tool community to seek
countermeasures, culminating now in an entire area of research and development:
network traffic obfuscation.  The goal is to force DPI-using
censors to abandon attempts at protocol identification, because accurate
identification at-scale requires prohibitive resources.  For example, by
developing tools whose traffic can reliably force DPI to misclassify the traffic
as that of a non-blacklisted tool.  It is network traffic obfuscation that we
will concern ourselves with in the remainder of this paper.

We want to emphasize that our focus on obfuscation does not belie the importance of
other aspects of censorship circumvention. IP- and DNS-based filtering are still
more prevalent than DPI-based mechanisms, and circumventing such filtering
remains a challenge in practice.  Ultimately circumvention requires
a variety of mechanisms that work in concert to defeat a censor and a system is
only as good as its weakest link.

\paragraph{Network traffic obfuscation threat-models. }
A \textit{threat model} is a particular combination of censor capabilities and
goals. We will provide only informal description of threat models
for the purposes of this article.
We emphasize that the threat models presented here
apply only to gathering information from network flows (IP information,
application-layer content, etc.), and not to censoring actions (blocking or
otherwise disrupting the communication) that this information enables.
Information may be gathered by using \emph{passive} or \emph{active} techniques.

A passive technique simply observes traffic flows.  One high-profile example of
passive techniques occurred during the February 2012 anniversary of the 1979
Iranian revolution. The Islamic Republic of Iran mounted a large-scale
censorship campaign, blocking various IP addresses associated with
anti-censorship tools such as torproject.org, and using DPI-based identification of TLS network flows.\footnote{See: \url{https://blog.torproject.org/blog/}\\\url{iran-partially-blocks-encrypted-network-traffic}}

An active technique modifies existing packet contents or injects completely new
packets. Examples of active attacks that manipulate connections by modifying
or dropping specific packets have appeared in the academic literature~\cite{parrot}. 
While far less common than passive techniques, active techniques have been used by the Chinese government in order to identify Tor proxies~\cite{winter2012great}.
We give a list of example passive and active attacks in \figref{fig:taxonomy}. For brevity, in the remainder of the paper we
focus primarily on passive attackers.

\begin{figure*}[t]
\center\footnotesize
\begin{tabular}{|l|l|p{4.25in}|c|c|}
\hline
\textbf{Country} & \textbf{Dates} & \textbf{Techniques and Targets} & \textbf{Passive/Active} \\
\hline
China  & 2007-- & IP/DNS blacklisting, \textbf{Tor TLS header fingerprinting, Tor-relay handshake probe} & Both \\
Iran & Feb.~2012 & IP blacklisting, \textbf{keyword flagging, TLS handshake fingerprinting} & Passive \\
Turkey & Early 2014  & IP/DNS blacklisting, BGP hijack of Twitter and Youtube & Both \\
Pakistan & Intermittent & IP/DNS blacklisting, \textbf{URL, keyword, and filetype flagging} & Passive \\
Syria & 2011-- & IP/DNS blacklisting, \textbf{instant message fingerprinting, social media keyword blacklisting} & Passive \\
Ethiopia & 2012 & \textbf{Skype detection, Tor TLS fingerprinting} & Passive \\
Belarus & 2014-- & IP/DNS blacklisting, \textbf{filtering according to packet content} & Passive \\
Kazakhstan & 2012 & \textbf{Encryption fingerprinting, Tor TLS handshake fingerprinting} & Passive  \\
Egypt & 2014-- & IP/DNS blacklisting, \textbf{instant message fingerprinting, social media keyword blacklisting} & Passive \\
\hline
\end{tabular}
\caption{Examples of censorship by nation states.  Techniques and targets in bold rely on DPI.}
\label{fig:taxonomy}
\end{figure*}

So far we have spoken mostly about the capabilities part of the threat model.
In terms of the goals, the typical assumption is that censors minimally want to
accurately classify traffic as either targeted (for application of policy) or
not.   Accuracy has two facets.  The first is a high true-positive rate, meaning
that the fraction of targeted traffic that is marked as such is close to one.
The second is a low false-positive rate, meaning that the fraction of non-target
traffic incorrectly marked as targeted is close to zero.    Obfuscation tools
are often said to be \emph{unobservable} if they can cause sufficiently low true-positive
and high false-positive rates.
What constitutes sufficient error rates, however, is largely
unknown. However, tools with the potential to induce
high false-positive rates are preferred as it is argued that censors will be less likely to block them.  Let us say more about why.

\paragraph{Sensitivity to collateral damage.}
Collateral damage occurs when a censor's detection mechanisms suffer from
false-positives, leading to mislabeling network flows as associated to a
blacklisted tool, protocol, or content when, in fact, the flow is not.  Why
should censors be sensitive to collateral damage? The answer is highly context
dependent, varies over time, and is ultimately a question of socio-political
factors. The example of the February 2012 censorship campaign by Iran is an
example of insensitive censorship (since the vast majority of TLS
traffic was almost certainly not used for circumvention or bound for hosts deemed offensive by
the regime). We note that this particularly broad censorship policy lasted for only
a short period of time, and that it is is typical for censorship policies to change,
in terms of targets and sensitivity, over time.  We will discuss later
how measuring sensitivity remains a critical open research challenge.

%

\section{Obfuscation Approaches}
\label{sec:systems}

Use of DPI by censors has inspired researchers and activists to explore a
variety of countermeasures aimed at obfuscating packet payloads.  We break down
approaches into four categories: encryption, randomization, mimicry, and
tunneling.

\paragraph{Encryption.} A natural first thought is that conventional encryption
is already a good obfuscator.  Indeed, encryption renders unintelligible the
data that is encrypted.  But most encryption protocols have plaintext headers in
each packet payload that identify the protocol \textit{itself}, and sometimes
indicate the particular application that is implementing the protocol.  We've
seen in Iran and China that plaintext headers contain fingerprintable parameters
that can help a censor distinguish between, say, Tor's implementation of TLS
from non-Tor implementations.  In short, just encrypting data does not guarantee
a level of unobservability sufficient for the goal of making DPI unreliable for
censors.

\paragraph{Randomization.} If conventional encryption, like plaintext protocols,
can leave exploitable fingerprints for DPI, why not attempt to randomize
\textit{every} byte in the packet payload? This strategy has been described as
attempting to make network flows ``look like
nothing'' by the obfs family of obfuscators~\cite{tor-pluggable-transports},
and is a strategy also used by Dust~\cite{dust} and ScrambleSuit~\cite{scramblesuit}.
ScrambleSuit and the obfs methods are designed specifically for the Tor
pluggable transport framework called ``Obfsproxy''.

The standard approach to randomizing a payload is to apply a stream cipher to
every byte.
Assuming a passive censor that does not have access to the stream
cipher secret key, this approach makes protocol identification difficult because
there are no fixed fingerprints to observe.  Because stream ciphers do not
provide the strong confidentiality and authenticity guarantees sought, they are
typically applied on top of modern encryption methods that do.

Underlying the use of randomizing, ``look-like-nothing'' obfuscators is the
assumption that censors use only blacklisting: they allow all traffic
except that which conforms to an explicit set of disallowed traffic types. If censors
use whitelisting, exclusively or in addition to blacklisting, then fully
randomized payloads will not avoid censorship.  This is by virtue of looking
like no actual network traffic.

\paragraph{Mimicry.}  To address whitelisting censors, a class of obfuscation methods attempt to
evade censorship by masquerading as a whitelisted protocol.  Loosely, rather
than trying to look like nothing, mimicry-based obfuscators try to make packet payloads look like something that they are not, but that will be identified by DPI as allowable.
A common example is to make payloads look like HTTP, which is rarely blocked because of its ubiquity.
Mimicry in the context of anti-censorship originates with the Tor project prepending HTTP headers to Tor packets~\cite{tor-pluggable-transports},
though mimicry can be thought of as steganography against a particular type of
adversary (DPI systems). Stegotorus~\cite{stegotorus} and
SkypeMorph~\cite{skypemorph}, for example, attempted to use steganographic
approaches to look like HTTP and Skype traffic, respectively. Unfortunately
they have prohibitively poor performance.

A minimalist but flexible approch to mimicry can be found in format-transforming
encryption (FTE)~\cite{usenix2014-fte}. As implemented
in Tor, uProxy and elsewhere, FTE allows programmers to use regular expressions
to specify the format of ciphertexts.  Because regular expressions are also used
by DPI, one can often precisely force misclassification of the protocol.

Mimicry obfuscators attempt to make packet payloads look
like a particular ``cover'' protocol, but do not actually use an implementation
of the targeted cover.  This means that the syntax and semantics of messages emanating from
mimicry systems may deviate, often significantly, from that of
messages conforming to the cover protocol.  As one example, a mimicking
obfuscator may produce a message purporting to be an {\tt HTTP GET} after having
just received a message purporting to be an {\tt HTTP GET}, thereby violating
proper HTTP semantics.

\paragraph{Tunneling.}
Instead of custom obfuscation
implementations of mimicry that reproduces some aspects of the cover protocol, tunneling
obfuscators transmit messages produced by an actual implementation of the cover
protocol.  Tunneling is of course not something new to obfuscation: all proxy
systems use tunneling over encrypted channels. So any VPN or HTTPS proxy might
be considered a tunneling obfuscator. The key difference is that tunneling
obfuscators are meant to obfuscate that an anti-censorship tool is being used,
and sometimes even obfuscate that encryption is being used. A VPN or proxy may
not (attempt to) hide such facts.

A notable example of a tunneling obfuscator is the Tor pluggable transport
Meek~\cite{fifield2015blocking}. Meek tunnels via TLS to HTTP load-balancers operated by
Google, Amazon CloudFront, or Microsoft Azure. This is referred to as ``domain
fronting'' because the ostensible domain revealed at the TCP layer (and so, in
the clear) is {\tt google.com} (say), but the underlying encrypted request is
for a domain that routes to a Meek-enabled Tor relay.\footnote{Interestingly,
a similar use of domain fronting was employed by the GoAgent system and was
popular in China in 2013.} As with Tor-TLS,  the
Meek obfuscator is at pains to ensure that tunneled Tor-connections to the
load balancer look like conventional TLS connections to the load balancer.
It's worth noting that
this approach relies on an undocumented ``feature'' of load-balancer
implementations, and it is conceivable that companies may modify
future load balancers in a way that prevents domain fronting.

The table appearing in \figref{fig:obfuscator-summary} summarizes a range of
obfuscators that have been deployed and compares them qualitatively in terms of
deployment, how keys are exchanged (should encryption be used to perform
obfuscation), and performance. It also includes a summary of attacks that have
been explored against them, as discussed in Sec.~\ref{sec:attacks}.

\begin{figure*}[t!]
\center \footnotesize
\begin{tabular}{|>{\raggedright}p{1in}|l|p{0.7in}|p{1in}|l|c|p{1.65in}|}
\hline
\textbf{Obfuscator} & \textbf{Type} & \textbf{Deployment} & \textbf{Key exchange} & \textbf{Perf.} & \textbf{Attacks} &  \textbf{Notes} \\
\hline
Cirripede~\cite{cirripede} &  tunneling & & In-band~(TLS) with pre-registration & High &  &Requires on-path cooperating ISP/routers\\[2pt]
CloudTransport~\cite{cloudtransport} & tunneling & & In-band (TLS) & Moderate & & Google or Amazon HTTPS, cloud storage as rendezvous \\[2pt]
Collage~\cite{burnett2010chipping} & mimicry & & n/a & Low & & Embed in user-generated content, use content servers as rendezvous\\[2pt]
Decoy Routing~\cite{decoyrouting} &  tunneling & & Out-of-band & High &  &Requires on-path cooperating routers\\[2pt]
Dust~\cite{dust} &  randomizer & uProxy& Out-of-band & High &  & Stream cipher w/ transmitted key material\\[2pt]
FTE~\cite{usenix2014-fte} & mimicry & Lantern, Tor, uProxy & Out-of-band & High & \cite{wang2015seeing}  & Flexible DPI-regex matching of HTTP, SSH, SMTP, etc. \\[2pt]
Flashproxy~\cite{fifield2012evading} & tunneling & Tor& n/a & Moderate  & &  Plaintext tunnel using websockets through browser-based proxies\\[2pt]
Infranet~\cite{feamster2002infranet} & mimicry & & In-band private & Low & &Modulate sequence of HTTP requests
to send data \\[2pt]
Marionette~\cite{usenix2015-marionette} & mimicry & & Out-of-band & Moderate  & &  Programmable mimicry including stateful and statistical properties\\[2pt]
Domain Fronting (e.g.~Meek~\cite{fifield2015blocking}) & tunneling & Tor, Psiphon, Lantern, uProxy, Psiphon, GoAgent & In-band (TLS) &  Moderate & \cite{wang2015seeing} & Google or Amazon HTTPS \\[2pt]
Non-PT Tor & tunneling &  Tor & In-band (TLS) & High & \cite{winter2012great} & Non-Tor TLS\\[2pt]
obfs\{2,3\}\cite{tor-pluggable-transports} & randomizer & Tor & Out-of-band & High & \cite{wang2015seeing} & Full encryption of messages\\[2pt]
obfs4~\cite{tor-pluggable-transports} & randomizer & Tor, Lantern & In-band private & High & \cite{wang2015seeing} & Uniform key exchange, full encryption of messages\\[2pt]
ScrambleSuit~\cite{scramblesuit} & randomizer & Tor & In-band private & High & \cite{wang2015seeing} & Uniform key exchange, full encryption of messages\\[2pt]
SkypeMorph~\cite{skypemorph} & mimicry & & Out-of-band (Skype) & Low & \cite{parrot} &  Mimics Skype connections \\[2pt]
Stegotorus~\cite{stegotorus} & mimicry & & In-band private & Low & \cite{parrot} & Mimics HTTP connections \\[2pt]
Telex~\cite{telex} &  tunneling & & In-band (modified TLS) & High &  &Requires on-path cooperating ISP/routers\\
\hline
\end{tabular}
\caption{Comparison of network obfuscation approaches.  Types correspond to our classifications of randomizer, mimicry, or tunneling.  Any notable deployments are listed.  The performance column (``Perf.'') is qualitative and based upon numbers reported in the original work.}
\label{fig:obfuscator-summary}
\end{figure*}

%

\section{Seeing Through Obfuscation}
\label{sec:attacks}

Can real censors detect obfuscators?  Researchers have begun preliminary
investigations into this question, and we will summarize their approaches here
focusing on passive techniques.
The takeaway message from what follows is that the answer is not yet clear.

\paragraph{Syntactic and semantic mismatch.} The first intensive investigation
into obfuscator detection is the work of Houmansadr, Brubaker, and
Shmatikov~\cite{parrot}.  They focus on mimicry systems in particular, and claim
that all such systems are fundamentally flawed because they will
never perfectly replicate all aspects of a target protocol implementation.

They look specifically at the extent to which messages emitted by currently implemented systems
abide by the proper syntax or semantics of the cover protocol.
By syntax we mean that the emitted messages are parsable as
cover protocol messages --- required headers are in the right places, the correct
character sets are used, etc.
Semantics refers instead to whether messages are meaningful relative to the
proper functioning of the protocol.

Most mimicry systems explicitly made no attempt to mimic \emph{all}\/ aspects of
the cover protocol, making the search for syntactic and semantic mismatches a
foregone conclusion. As examples, Stegotorus doesn't implement most of HTTP, and
the packets sent when using FTE (in a way that favors performance) contains
large numbers of random, unformatted bits.  The authors show how to take
advantage of such mismatches. In particular, they detail a number of ways to
detect Stegotorus, SkypeMorph, and CensorSpoofer.  None of these obfuscators are
actually in use (although for reasons unrelated to security), and so these
attacks were only tested against simulated stand-ins for deployed systems. It is
possible that real deployments would be more (or less!) vulnerable to their
attack methods.

We also note that the authors did not measure the collateral damage of their tests, i.e.,
to what extent the attacks they develop would misclassify benign cover-protocol
traffic as obfuscated. Subsequent work~\cite{wang2015seeing} showed
that some of their tests would lead to a false positive rate of 37\% if
naively deployed, while others appear effective at avoiding collateral damage.

\paragraph{Entropy tests.}  The semantic and syntactic checks above will not
help one identify the use of randomizing obfuscators, as these explicitly try to look like no actual protocol.
Wang et al.~\cite{wang2015seeing} evaluated the use of efficient entropy tests
in the context of detection of randomizing obfuscators.
Their techniques focus on a small window of bytes occurring in the first few messages of an observed flow.
The intuition for why this works is that the packets emitted by all
non-obfuscated protocols always contain some plaintext headers, whereas
randomizers by definition have random data in these locations.  They showed
their test against obfsproxy achieved a relatively low false-positive rate
of only 0.2\% when used against non-obfuscated network traces collected from a
university campus. Whether or not this constitutes a sufficiently small false-positive rate in practice is both context-dependent and unclear.

\paragraph{Traffic analysis.} Even without parsing packet payloads to exploit the actual byte values being carried,
obfuscator-produced traffic may exhibit packet lengths, inter-packet timings
and directional traffic patterns that deviate from the background network traffic.
Using such meta-data to infer information about communications is referred to as traffic analysis.

One example of deviant traffic behavior arises in Meek.
Conventionally, HTTPS
connections involve two message flows, one smaller upstream request (client to
server), and then a longer downstream message containing the requested content.
With Meek, however, the client must frequently send requests to see if new data
is available. To a network DPI system, this may stand out as an
uncharacteristically large number of upstream messages.  The ability to
exploit this by a DPI was investigated by Wang et al., who gave a
machine-learning-based approach that detected Meek with true positive rate of
0.98 and false positive rate of
0.0002 or less. But their
detector may not be easy to use in practice
as it only worked well when trained and tested on traces from the same
network environment.

For fear of traffic analysis, some obfuscators include mechanisms to try to
randomize meta-data such as packet lengths and packet inter-arrival time. Such
countermeasures borrow techniques from the extensive literature on
website fingerprinting and its countermeasures~\cite{juarez2014critical}.
How effective these countermeasures are in the obfuscation context is unknown.

%

\section{Challenges and Open Questions}
\label{sec:challenges}

So far, we have provided an overview of the rapid evolution of
censorship and censorship-evasion technologies in the context of network traffic
obfuscation.  Now we turn to the many open questions and challenges
that remain.  We cluster them into three general areas:  understanding users,
understanding censors, and future-proofing obfuscation systems.  Within
each of these areas, we call out specific tasks that need addressing.

\subsection{Understanding Users}

Beyond anecdotes and folklore, very little is understood
about the users of censorship-circumvention tools, and this hamstrings the
development of effective tools. Research that takes a
principled approach to understanding  users and their situations could lead to
significant improvements. But performing user studies in this
context is particularly challenging.  The use of censorship-circumvention tools is illegal in some places
and, in extreme cases, exposing users could put them in physical danger.

\paragraph{Develop guidelines for user studies.}
The research community might produce guidelines for performing ethical human subjects
research in this space.\footnote{Standard IRB-style human subjects reviews are
also important, but could be complemented by further
community-driven guidelines.} These guidelines would do well to account for regional and cultural
differences, reflect the challenge of obtaining \emph{informed} consent, and
perform measurement in a transparent yet privacy-sensitive way.

\paragraph{Perform user studies.}
The guidelines could be put to use in the design of ethical research studies.
Key questions that might
be addressed include: What makes users decide to use a system, or not? How do
users first learn of the availability of tools and then obtain them? What are
user perceptions regarding risk, ethics, and legality of tool use? How do users
react to different kinds of performance or functionality loss when using
circumvention tools? How important is it to users that they be able to
convincingly deny that they used a tool?

\subsection{Understanding Censors}

Perhaps the most important theme emerging from
this paper is that researchers do not yet have a clear understanding of the
adversary.  This prevents us from knowing whether or not a
proposed obfuscation method is truly  ``broken'' or targets an appropriate design trade-offs
between security, performance, and usability.

\paragraph{Characterize ``normal'' traffic.} Circumvention technologies depend on sensitivity to
collateral damage, but at present we lack models for how normal, unobfuscated
traffic appears to DPI systems. This is especially important for mimicry-based
obfuscators which, by design, try to avoid detection by
appearing to be normal traffic.

So far, researchers have either generated synthetic data
sets or, in one case~\cite{wang2015seeing}, gathered traces from a university's
networks. The problem with synthetic traces, and even university network captures, is that they are unlikely to
properly represent the wide diversity of traffic that censors observe.

One approach here would be to develop partnerships with large network
operators (e.g.\  ISPs) in order to perform studies. We note that for obfuscation research, one
unfortunately needs what's referred to as full-packet capture, i.e.\ the entire
packet contents, including application-layer data.  This could have
significant privacy implications for the users of the network, so
special care is demanded.  Perhaps researchers could develop
new techniques for doing studies without ever storing packets, and/or
incorporate other privacy mechanisms (c.f.,~\cite{miklas2009bunker}).
This would
improve the privacy of such studies and, one hopes, reduce reluctance of
network operators to help perform them.

One powerful artifact of such studies would be useful models of
``normal'' traffic in various settings.  These could be used to build
more accurate simulations to support further analysis and tool
testing.  What would make a model ``useful'' is not currently well
defined.

\paragraph{Models of collateral damage.} It seems clear that our
threat models must account for the sensitivity of censors to collateral damage.
Underlying all censorship circumvention tools is the assumption that censors want to limit collateral damage, at least to some
extent. There are a number of open questions here.

First is characterizing what types of collateral damage are of concern to
censors.  We expect that tolerance to false-positives varies from country to country, and across time within a single
country. Moreover, not all collateral damage is the same: it may be that
causing collateral damage of certain forms is more costly to censors in terms of political or economic
repercussions. Domain fronting systems, such as
CloudTransport and Meek both rely on the hope that
blocking popular cloud services, e.g. those provided by Google or
Amazon, is untenable. However, some countries like China \emph{have} blocked such services
entirely for periods of time. A methodology for systems to reason about
collateral damage is clearly needed.

The challenge is for models to incorporate parameters in a
way that captures informal statements of the form ``the censor will not tolerate
more than an $n$\% false-positive rate.''  Few existing works attempt to
measure false-positive rates in any sort of methodical manner.  Absent of such
attempts, we have no way to state that a given technique will induce collateral
damage that is intolerable to the modeled censor.

A starting point would be to build threat models parameterized by false-positive
rates and standardized methods to identify network traffic.  Researchers could
then present receiver operating characteristic (ROC) curves with their approach.
ROC curves and measures such as area-under-the-curve (AUC) are already standard
practice in machine learning, and we recommend bringing these forms of analysis
to circumvention technology development.

\paragraph{Accessible censorship adversary labs.} Given the widespread and
increasing use of commercial DPI devices, good threat models should be informed
by what devices can reasonably be expected to do now and in the near future ---
what computational tasks can they perform (regular expression matching? full
parsing of packet payloads? intensive statistical tests?), how much state can
they keep (a single packet? all packets in a TCP connection? all connections
from a particular IP address?). This assessment will be context-dependent: it
will be impacted by the volume of traffic and the assumed sensitivity to
collateral damage.  This suggest that a hierarchy of threat models should be
developed to guide building useful systems, rather than attempting to get
consensus on a single model.

A hurdle has been obtaining
access to the kinds of equipment used by censors. DPI systems are
expensive proprietary devices, and they come with restrictive licensing agreements
that typically prevent any benchmarking. This means that researchers,
even if they purchase one, may be contractually obligated not to perform
research with them. While hardware itself can often be purchased second-hand on
sites like Ebay, they may have out-of-date software and not fully reflect
the abilities of fully updated deployments.

The community would benefit from a concerted effort to overcome these hurdles,
with the goal of building a widely usable censorship lab.  The lab would need to
obtain realistic, preferably state of the art, devices with realistic
configurations.  Less restrictive licensing would need to be negotiated.  We
might look to previous scientific infrastructure projects for inspiration, such
as PlanetLab, CloudLab, and the National Cyber Range.  We envision such a lab
acting as a repository for reference datasets of ``normal'' traffic (as
discussed above), in addition to well providing a battery of existing censorship
tools.  Ideally, the lab would support remote access for researchers.

\subsection{Future-proofing Obfuscation Systems}

We expect that user studies and better understanding of censors will yield
significant insights into how to build better circumvention systems.
Eventually our goal should be to adapt to, or even defuse, the current
armsrace, and give to developers the techniques needed to have future-proof
obfuscation mechanisms.  By this we mean mechanisms that can maintain
connectivity for users even in the face of adaptive censors that know the design
of the system.
We now summarize the research directions needed to achieve this.

\paragraph{Develop formal foundations for obfuscation research.} Currently there
is a dearth of formal abstractions --- formal threat models in particular --- upon
which to lay principled foundations for obfuscation.  One approach is to follow
the lead of modern cryptographic theory, where giving precise adversarial
models, security goals, and syntax for primitives are primary research
considerations. This approach enables clear scientific discussion and objective
comparison of systems relative to specified adversarial models.  In short, good
models are a significant aid in the development of new tools.

Unlike traditional cryptographic problems, where efficient constructions have
been shown to meet very pessimistic adversarial models, the
censorship-circumvention setting may be one that benefits from more realistic
models, informed and evolved by experimental evidence and measurements from the
field.

\paragraph{Build general frameworks.}  Obfuscation mechanisms have mostly been built
with a particular target circumvention system in mind.
But obfuscation methods designed for a specific circumvention system may not be easily adapted for another.
For example, Tor is TCP-based, so Tor-oriented obfuscation methods need not consider unreliable transports, but I2P and uProxy must.

To have broader impact, the obfuscation community would benefit from general libraries and frameworks.
In addition to cutting down on aggregate implementation effort, having broadly useful obfuscation libraries could
facilitate rapid rollouts of countermeasures to new attacks --- not every tool
would need to have a different patch developed.  Of course, building more-general libraries will require
community consensus on typical architectures, APIs, and protocols for
negotiating which obfuscation to use.

\paragraph{Increase the use of encryption.} Recent efforts to expand the
use of standard encryption tools like TLS make it harder to perform
censorship. If most of the Internet is end-to-end encrypted using
TLS, then censors are harder pressed to use DPI to detect content they wish to block.
Indeed we view improved censorship circumvention as a nice ancillary benefit to the already
laudable goal of ubiquitous encryption on the Internet.

However, recall our earlier warning that encryption is not a silver bullet for
censorship circumvention. Particular applications that use common encryption
protocols can still be fingerprinted, often by information sent in the
clear 
such as headers, packet sizes, and timing.

Censors are also sometimes positioned to break end-to-end security,
performing man-in-the-middle (MitM) attacks either by directly abusing the web's public-key
infrastructure (as occurred in Turkey in 2012)\footnote{\url{https://googleonlinesecurity.blogspot.com/2013/01/enhancing-digital-certificate-security.html}} or by forcing clients to install
root certificates that trick systems into allowing MitM proxies. The latter is
common in corporate settings already, and retooling those solutions to
nation-state levels is part of the threat landscape that should be considered.

Simultaneously, encryption standards could do more to support obfuscation.
For example, they could encrypt more of the handshake required during connection establishment. Ideally all bits emitted
would be completely random-looking to the censor.  (This is targeted by the obfs4 system.)
If all TLS connections achieved this then the entropy-based attacks discussed in Section~\ref{sec:attacks} would no longer be
effective.



%

\section{Conclusion}
\label{sec:futurework}

Censorship-circumvention technology is a relatively young field of research but
already has a wide diversity of approaches and systems.   This suggests the need
to reflect and consolidate what has been learned and to propose open questions
to facilitate the field's development.

Perhaps the most striking observation is that \emph{researchers do not yet have
a clear and common model of the adversary}. We also noted significant
challenges to understanding users and making obfuscation systems robust
to future attacks. These limitations are preventing researchers from knowing
whether or not a proposed obfuscation method is vulnerable in practice (verses
just in theory); and whether a design has made appropriate trade-offs between
security, performance, ease of use, or if it is failing to address user needs.

These are open questions in large part due to a surfeit of methodological
challenges faced by researchers in this area.  Chief among these are the
difficulty of obtaining access to network traffic without violating privacy, of
procuring DPI systems used by censors, and of performing user studies within
relevant populations.  Such challenges make principled progress difficult,
but also offer
opportunities for creative workarounds.
We suggested some specific tasks and research directions that the
community might take up in order to make progress, including developing
guidelines for user studies and building a shared lab for evaluating DPI
capabilities.

Unfortunately Internet censorship will be around for the foreseeable future.
Researchers, activists, and governments supporting the right to unfettered
information online should continue to work towards a future in which
network traffic obfuscation prevails in the face of sophisticated,
well-provisioned censors. This will provide a critical piece to the
techno-social puzzle enabling people around the world access to information.

\section*{Acknowledgements}

The authors would like to thank the participants of the 2014 Obfuscation
workshop organized by Google for their valuable contributions.

\bibliographystyle{abbrv}
\bibliography{bib}

\end{document}